\documentstyle[titlepage,12pt] {article}
\title{
A quantitative measurement of spatial order in ventricular fibrillation}
\author{P.V. Bayly \and E.E. Johnson \and P.D. Wolf
         \and H.S. Greenside \and W.M. Smith \and R.E. Ideker           \\
                                                      \\
{\em        From the Engineering Research Center                       }\\
{\em        for Emerging Cardiovascular Technology                     }\\
{\em        School of Engineering                                      }\\
{\em        Duke University                                            }\\
{\em        Durham, North Carolina 27708                               }\\
{\em        and the Departments of Pathology and Medicine              }\\
{\em        Duke University Medical Center                             }\\
{\em        Durham, North Carolina 27710                               }\\
                                                                        \\
{\em        Supported in part by the National Institutes of Health     }\\
{\em        research grants HL-28429, HL-44066, and HL-33637           }\\
{\em        and the National Science Foundation                        }\\
{\em        Engineering Research Center Grant CDR-8622201              }\\
}
\begin{document}
\maketitle
\pagestyle{plain}                      % puts in page numbers only
\pagenumbering{arabic}

\pagebreak
%\begin{abstract}
\begin{centering}
\section*{ {\large \bf Abstract} }
\end{centering}

\noindent
{\it Introduction:}
The degree of organization in ventricular fibrillation (VF) is not known.
As an objective measurement of spatial order, spatial
correlation functions and their characteristic
lengths were estimated from
epicardial electrograms of pigs in VF.

\noindent
{\it Methods:}
VF was
induced by premature stimulation in 5 pigs.
Electrograms were
simultaneously recorded with a $22 \times 23$ array of unipolar electrodes
spaced 1.12 mm apart.
Data were obtained by sampling the signals at 2000 Hz for 20 minutes
immediately
after the initiation of VF. Correlations between all pairs of signals were
computed
at various times.
Correlation lengths were estimated from
the decay of average correlation as a function of electrode separation.

\noindent
{\it Results:}
The correlation length of the VF in pigs was found to be
approximately 4-10 mm,
varying as fibrillation progressed.
The degree of correlation decreased in the first 4 seconds after fibrillation
then increased over the next minute.

\noindent
{\it Conclusions:}
The correlation length is much
smaller than the scale of the heart, suggesting that many
independent regions of activity exist on the epicardium at any one time.
On the other hand, the correlation length is 4 to 10 times the
interelectrode spacing,
indicating that some coherence is present.
These results
imply that the heart behaves during VF as a high-dimensional, but not random,
system involving many spatial degrees of freedom,
which may explain the lack of convergence of fractal
dimension estimates reported in the literature.
Changes in the correlation length also suggest that
VF reorganizes slightly in the first minute
after an initial breakdown in structure.
\newline
{\bf Key words:} spatial correlation, correlation length,
organization, dynamics, ventricular fibrillation, sudden death.

%\end{abstract}
\pagebreak
\section{Introduction}

The ventricular myocardium is a nonlinear,
excitable system which can appear almost totally synchronized
in sinus rhythm, or completely disorganized in fibrillation.
Fifty years ago Wiggers described four stages of
ventricular fibrillation;
his classification relies strongly on qualitative descriptions of the
degree of spatial organization such as ``undulatory
contractions'' (stage 1), ``convulsive incoordination'' (stage 2),
``tremulous incoordination'' (stage 3), and ``feeble contraction
wavelets'' (stage 4)
\cite{wiggers}.
Reentrant spatial structures such as wavelets or rotors are
thought to provide
the underlying mechanism of VF. Rotating activation patterns have been
observed in experiments in animals and
in isolated myocardium
\cite{frazier,allessie,ideker90,davidenko};
they have also been widely studied in mathematical models of excitable media
\cite{barkley,keener,gerhardt}.
In general, greater spatial order is associated with fewer, larger structures.
In this paper we explore an objective measure of the {\em amount}
of spatial organization: the spatial correlation length.

Many investigators
have examined the organization of activation in the heart
by detecting potentials at a number of
locations in the myocardium with an array of spatially
distributed electrodes
\cite{allessie,ideker81,elsherif}.
Organization and spatial structure have been documented by
following the progress of individual waves of activation.
Isochronal maps, which
depict the position of the wave front as time progresses, or
isopotential maps, which show lines of constant voltage on the surface of the
heart at a single instant, illustrate the size and type of waveforms which
occur during VF. Isochronal maps are more commonly used than
isopotential plots, since one map
can show the position of the wave front at many instants of time.
However, objective, statistical descriptions of activity are also needed
to compare with results from simulations and from other
experiments.

The question of spatial organization is related to the widely speculated
relationship between nonlinear dynamics, chaos, and fibrillation
\cite{goldbergerphyd,kaplan+cohen,guevara+glass}.
The mathematical term ``chaos'' refers specifically to the temporally
aperiodic, unpredictable behavior of some nonlinear deterministic systems
characterized
by broad-band power spectra and extreme sensitivity to initial conditions.
A necessary condition for chaos is at least one positive Lyapunov exponent,
which
describes the exponential separation of nearby orbits.
The concepts of strange attractors and chaos have illuminated
the behavior of simple, few degree-of-freedom, nonlinear
systems.
The methods of low-dimensional chaos also apply to the nonlinear
dynamics of
multi-degree-of-freedom systems which
involve very few modes
\cite{holmesamr90}
such as an oscillating fluid system
\cite{swinney+gollub}.

Temporally complicated, but spatially organized, behavior
of  extended systems may often  be well
modeled by low-dimensional chaos.
In these cases one may obtain useful quantities,
like Lyapunov exponents, and analyze simplified models
which retain the basic mechanisms of instability and disorder.
Spatially incoherent dynamics in an extended system are
inherently high-dimensional, meaning that
many independent variables are required to describe its state.
These cases are impractical to treat as chaotic
\cite{guckenheimer}.
Because of this difference,
determining the degree of spatial organization
of VF is important in developing useful models of fibrillation
and finding effective clinical countermeasures.

In other fields, notably fluid mechanics, a
numerical measure of the characteristic size of a spatial process has
frequently been extracted:
the correlation length. There is more than one definition of the
correlation length; the quantity, $\lambda$, used here
describes the rate of
decay of the spatial correlation function as a function of
distance. It has been calculated for many experiments,
and has been interpreted as the characteristic length scale of the typical
spatial structure in the experiment: the diameter of the average
wavelet, eddy, or rotor, for example
\cite{panchev,lumley,tufillaro,hohenberg+shraiman,coullet}.

While VF has been the subject of much study and more speculation,
an objective, quantitative description of its spatial
organization has not been put forward. Clear experimental
results are needed to compare with the burgeoning supply
of data from numerical models
\cite{panfilov,markus}.
In this study, the correlation length of ventricular fibrillation has been
calculated for several intervals, at various times after the initiation
of VF, in each of five pig hearts. The application of the method and the
experimental procedure are explained in more
detail below.

\section{Methods}
\subsection{Spatial correlation function}

The spatial correlation function, $R(d {\bf x})$, is
a scalar function describing the
average correlation
between all pairs of measurements separated by the vector $d {\bf x}$.
For a general, real-valued function of time and space, $g(t,{\bf x})$,
the spatial correlation function is
\cite{pratt}:

\begin{equation}
R(d {\bf x}) = E \{ \, g(t,{\bf x}) \, g(t,{\bf x} + d {\bf x}) \},
\end{equation}

\noindent
where $E$ denotes an ensemble average.
For a stationary, ergodic process, the value of the spatial
correlation function is obtained from the following integral.
Integration is over
the time interval $-T$ to $T$ and over the spatial domain $S$.

\begin{equation}
R(d {\bf x}) =  \lim_{T \rightarrow \infty}
                     \frac {1}{2T}
                     \int_{S}{
                     \int_{-T}^{T}
                     {g(t,{\bf x}) \, g(t,{\bf x} + d {\bf x}) \, dt}
		     \, d{\bf x}}.
\end{equation}

\noindent
The spatial correlation function
is completely analogous to the commonly encountered
temporal autocorrelation function,
\begin{equation}
 A(\tau) = \lim_{T \rightarrow \infty}
            \frac {1}{2T}  \int_{-T}^{T}{f(t) \, f(t + \tau) \, dt} .
\end{equation}
Sampling the continuous process, $g(t,{\bf x})$,
with a 2-dimensional array of equally spaced sensors leads to a
set of discrete measurements.  If P time samples are taken
from each sensor of an $M \times N$ array, one obtains the values:

\begin{equation}
c_{ijk} = g(t_{k},{\bf x}_{ij}), \, i=1,2,...,M, \, j=1,2,...,N, \,
k=1,2,...,P,
\end{equation}

\noindent
where $t_{k}$ is the time of the $k^{th}$ sample and ${\bf x}_{ij}$ is
the location of the sensor in the $i^{th}$ row and $j^{th}$ column
of the sensor array. For a regularly spaced array with a
separation of $h$ units between adjacent sensors,
${\bf x}_{ij} = [h \, (i-1), h \, (j-1)]^{T}$, and
$d {\bf x} = [h \, \Delta i, h \, \Delta j ]^{T}$.

The spatial correlation function of this discrete data set is
ideally approximated by the sum:

\begin{equation}
R(\Delta{i}, \Delta{j}) = \frac {1}{Q(\Delta{i}, \Delta{j})}
                          \sum_{i=1}^{M} \sum_{j=1}^{N} \sum_{k=1}^{P}
                          c_{ijk} c_{(i + \Delta i)(j + \Delta j) k}.
\end{equation}

\noindent
Here $Q(\Delta{i}, \Delta{j})$ is the total number of
simultaneous samples which are separated in space by the
the vector $[\Delta{i}, \Delta{j}]^{T}$.
The quantitities $\Delta i$ and $\Delta j$ may vary from
$-(M-1)$ to $(M-1)$ and
from $-(N-1)$ to $(N-1)$, respectively.

The resulting
estimates of spatial correlation are only statistically
valid for values of
$\Delta i$ and $\Delta j$ which allow an adequate number of
correlations to be averaged; for example, in an
$M \times N$ array there are only two sensors
separated by a displacement of $[(M-1), (N-1)]^{T}$, but
$MN$ sensors separated by the displacement $[0,0]^{T}$.
Accordingly, correlations will not be presented for separations
larger than 3/4 of the shorter side of the rectangular
array.

Just as the temporal autocorrelation function of a harmonic
signal is a sinusoidal curve with the same period, the spatial
correlation function
of a travelling plane wave is a
sinusoidal surface with the same wavelength (Figure 1). The spatial
correlation function of uncorrelated random noise
is close to zero everywhere except at the origin (Figure 2).
In many real,
physical systems (turbulent fluids, capillary waves),
the envelope of the correlation function does not
remain constant, as in the perfectly ordered plane wave, nor does it
vanish as in the case of the random noise. Instead, it usually decays
at some characteristic rate. For small ${\vline d {\bf x} \vline}$
the decay can be assumed to be exponential, as:
\begin{equation}
R(\vline d {\bf x} \vline) \propto
e^{ - \vline d {\bf x} \vline / \lambda }.
\end{equation}

To illustrate how we extract the correlation length,
Figure 3a shows a typical decaying correlation, $R$, plotted as a
function of the vector separation between sensors,
$[dx, dy]^{T}$.
The solid curve of Figure 3b is obtained by averaging
the surface of Figure 3a over all angles to obtain a plot of $R$ vs
$\vline d {\bf x} \vline$.
The decay of the correlation curve is
described by the parameter, $\lambda$, obtained from a
least-squares exponential curve
fit also shown in the Figure 3b.
Two numbers can be extracted from the original curve of
correlation vs separation:
$L_{e}$ is the length needed for the correlation to decay to $1/e$,
and $L_{0}$ is the value of $\vline d {\bf x} \vline$
at which the correlation first drops to zero.
$L_{e}$  may be used as a check on the value of $\lambda$.

Although the exponential curve is not always the optimal fit, it
is a simple, straightforward, repeatable, and widely
used approach. Values of $\lambda$  vary quite
closely with
values of $L_{e}$ (see Figure 3 and Table 1).
It is impractical to use $L_{e}$ or $L_{0}$ if not all curves decay
to $1/e$ or 0.

A wide variety of processes have been characterized
by their correlation lengths.
Hohenberg and Shraiman in reference
\cite{hohenberg+shraiman}
used the correlation
length, $ \lambda $, to describe the degree of
coherence of one of the simplest numerical models of
disorder in a spatially extended
nonlinear system: the
1-dimensional Kuramoto-Sivashinsky equation.
Tufillaro and colleagues in
\cite{tufillaro} studied the correlation length of a system
of small surface waves in water as a forcing parameter was varied.
Coullet and coworkers in
\cite{coullet}, recently used $\lambda$
as a quantitative measure of spatial organization in
numerical studies of
the complex Ginzburg-Landau equation,
a 2-D system which exhibits spiral patterns.
They show that for their model, the spatial
correlation length corresponds specifically to the mean distance between
spiral ``defects'', which is
presumably also the average diameter of the structures
\cite{coullet}.
Although no systematic evidence is presented of such a specific meaning
for the correlation
length of fibrillation, we interpret it
as roughly the average length
of a patch of coherent activity.

\subsection{Experiment}

Fibrillation data were obtained from 5 anesthetized pigs, each weighing
approximately 25-27 kg.
The animals were anesthetized with pentothal 5\%, and respiration
was supported by a mechanical ventilator. Anesthesia and fluid levels
were maintained with an intravenous line. A line from the femoral
artery allowed the continuous monitoring of hemodynamic status and blood gas
values. The animals received doses of the
drugs ketamine and acepromazine before anesthesia was started,
and intravenous succinylcholine was used to induce and maintain
skeletal muscle paralysis. Normal blood gas, electrolyte, and body
temperature levels were maintained.

The chest was opened using a median sternotomy, and the heart was
exposed and suspended in a pericardial cradle. The electrode plaque
was sutured to the epicardium of the right or left ventricle.
The electrode plaque is a plastic block containing 524 Ag-AgCl
electrodes, 506 of which form a $22 \times 23$ array. The electrodes in the
grid are spaced 1.12 mm apart, providing total dimensions
of $2.35 \times 2.46$ cm. A photograph of the array is
provided in Figure 4. The electrode spacing was chosen to avoid
significant spatial aliasing while allowing the plaque to cover as
large an area as possible.
A previous study has provided evidence that very little spatial aliasing error
is introduced with 1.12 mm spacing \cite{baylycinc}.

VF was induced by applying a premature
stimulus to the ventricular muscle. In this
procedure the muscle was paced with relatively small, regular,
current stimuli (S1 stimuli) applied to the myocardium through
a stainless steel electrode. The stimulus electrode was sutured to
the epicardium
about 1 cm from the electrode array.
The S1 stimuli were spaced 350 ms apart and each pulse delivered current
equal to twice the diastolic theshold.
After 10 equally spaced S1
pulses were applied, a larger (50 mA) current stimulus, S2,
was applied through another stainless steel electrode located
1 to 2 cm away.
The S2 stimulus was applied prematurely, while
some of the tissue was still in an inexcitable state from the
previous S1 beat.
The large, premature stimulus
is believed to induce rotors as it intersects the partially refractory region
\cite{frazier}.
This method is somewhat analogous to the procedure used by Davidenko
and colleagues to initiate rotating waves in isolated sheep myocardium
\cite{davidenko},
and to the initial conditions used by Barkley and others in numerical
models of other excitable media
\cite{barkley}.
Reference
\cite{johnsoncirc}
describes the experiment in more technical detail.

Simultaneous time series of the electrical potential at each of the
electrodes were recorded with 14-bit
precision at a sampling rate of 2000 Hz, using a
528-channel mapping system
\cite{patwolf}. The signals
were prefiltered with a 500 Hz low-pass filter to prevent temporal aliasing and
with a 0.5 Hz high-pass filter to eliminate low-frequency artifact.
These frequency cutoffs are well removed from the characteristic
frequency of VF in pigs, which is usually near 10 Hz
\cite{johnsonjacc}. Potentials were measured relative to a reference voltage
from the animal's left leg.

For each experiment, 20 minutes of
data were recorded immediately after VF was induced.
Subsets of these data were taken from the first few seconds
of fibrillation, and
after 15 seconds, 30 seconds, 1 minute, 2 minutes,
5 minutes, and 10 minutes of VF. Several spatial correlation functions
were calculated using one-second
blocks of data taken from each
time interval. The one-second block size was chosen to ensure
stationarity in time while including several cycles of
activation;
in several test cases the time interval was varied from 0.1 to 1.0
seconds, with only
slight quantitative effects.

\section{Results}

A typical time series recorded at a single electrode just after
the initiation of
fibrillation  is shown in Figure 5.
Figure 6 displays a sequence of spatial patterns recorded by the
electrode grid in the first second after the initiation of VF. The spatial
oscillations do not appear to be smaller than the size of the array.
Two seconds later the structure seems to have broken up, as the
sequence in Figure 7 indicates. By 60 seconds, however,
larger structures have reappeared (Figure 8).

Spatial correlation estimates track these changes in organization
consistently. Figure 9 shows
the spatial correlation computed from one-second blocks of data
in the first, second, third and sixty-first seconds after the initiation of VF
in one heart (experiment \#3).
In the first second the correlation surface falls off
relatively slowly from the peak at the origin (Figure 9a).
As disorganization
increases (more structures of smaller size) the
fall-off becomes more rapid (Figures 9b, 9c) and the surfaces approach the
delta function of Figure 2. However, at 60 seconds
the fall-off is noticeably less sharp (Figure 9d).

To extract a correlation length from the surface
of spatial correlation
the correlation estimate was averaged over all directions.
Curves of average correlation vs. magnitude of
separation are shown in Figure 10; these curves were
obtained by averaging
the spatial correlation surfaces of Figure 9.

The curves in Figure 10 decay roughly exponentially, and were fitted
using a least squares algorithm. The constant, $\lambda$,
for each curve is tabulated
in Table 1. The length needed for the
correlation to decay to a
value of $1/e$, $L_{e}$, and the value of the first zero-crossing
of the curve, $L_{0}$, are also presented. It is apparent
that $\lambda$, $L_{e}$ and $L_{0}$ vary together.
We can also see
that the qualitatively more organized behavior has a longer
correlation length.

\begin{table}[tbph]
\begin{center}

\begin{tabular}{||r|r|r|r|r|r||}  \hline
Time     & $\lambda$  & $L_{e}$ & $L_{0}$ \\
\hline
 1 sec   &    11.4       &  10.2     &  $> 15.0$      \\
 2 sec   &    6.3       &  6.0       &  10.8    \\
 3 sec   &    5.0       &  4.6       &  9.5     \\
 60 sec  &    8.7       &  7.9       &  12.7      \\
\hline
\end{tabular}
\end{center}
\caption{Characteristic lengths (in mm) obtained from the correlation
		curves of Figure 9 (experiment \#3): the correlation length or
		constant of exponential spatial decay,
                $\lambda$; the length to decay to 1/e, $L_{e}$;
                and the first zero-crossing, $L_{0}$.}
\label{tab:values1}
\end{table}

Figures 11 and 12 show spatial correlation
vs. vector separation and
spatial correlation vs. the magnitude of separation
for an episode of fibrillation in a different heart
(experiment \#4).
Although the progress of the arrhythmia is different,
the characteristic correlation curves of Figure 12 are
qualitatively similar to those
in Figure 10. The differences in the correlation length
capture the quantitative
differences in spatial organization.

Similar analyses were performed for each of the five experiments
and correlation lengths were calculated for data at various
times during VF. Table 2
provides a summary of the correlation lengths calculated
at several different times, for each experiment.
The correlation lengths fall mostly between 4 mm
and 10 mm.

% Table of lengths of all experiments
\begin{table}[tbph]
\begin{center}

\begin{tabular}{||r|r|r|r|r|r||}  \hline
Time     & Exp \#1   & Exp \#2  & Exp \#3  & Exp \#4  &Exp \#5 \\
\hline
 1 sec   & 11.3  & 5.9   & 11.4   & 6.1   &  3.8  \\
 3 sec   &  4.4  & 3.3   & 5.0   & 5.2   &  3.7  \\
 15 sec  &  4.8  & 3.7   & 6.0   & 6.0   &  4.1  \\
 30 sec  &  4.8  & 4.8   & 7.6   & 4.7   &  3.7  \\
 60 sec  &  5.0  & 5.4   & 8.7   & 8.1   &  5.0  \\
120 sec  &  6.0  & 5.3   & 6.7   & 8.4   &  4.6  \\
300 sec  &  5.4  & 4.8   & 5.3   & 6.7   &  4.2  \\
600 sec  &  5.7  & 4.9   & 5.7   & 6.4   &  4.1  \\
\hline
\end{tabular}
\end{center}

\caption{Correlation lengths, $\lambda$, (in mm) calculated from
         VF data during the progression of fibrillation
         in  5 experiments. Each value was calculated from
	 a one-second block of data recorded at the indicated
	 time after the initiation of VF.}
\label{tab:values2}
\end{table}

In this study it was found that the correlation length of VF decreased
over the first few seconds
but often increased as fibrillation progressed; in each experiment
the correlation length after 3 seconds was
shorter than both the correlation length after 1 second and
the correlation length after 60 seconds. These
differences were statistically significant with $p<0.05$. The mean correlation
lengths (over the 5 experiments) as VF progressed are shown in Figure 13.

\section{Discussion}

\subsection{Interpretation of results}

The correlation length is a widely used measurement of spatial
organization which can be used to estimate the degree
of spatial organization of physical processes. Longer correlation
lengths are associated with larger structures and
apparently more organized behavior.
This study suggests
that correlation lengths may be extracted from experimental epicardial
electrograms and that those lengths may be used to compare different
stages and types of ventricular fibrillation.

Moe and his co-workers postulated in 1964 that the size and number
of wavelets in the ``wandering wavelet'' model of ventricular
fibrillation was critical
to the development of an episode of fibrillation
\cite{moe}.
The broad interpretation of spatial correlation
is not
restricted to a certain physical process, nor does it explicitly
provide evidence for a particular mechanism (such as
wandering wavelets, or spiral waves). The physical mechanism should, however,
be consistent with the specific numerical
values of the correlation length.

The initial breakup of organization during fibrillation was first
 observed years ago. It is described
by Wiggers as the transition between ``undulatory'' motion and
``convulsive incoordination''
\cite{wiggers}.
The apparent reorganization of VF after 30-60 seconds is not well known,
but is consistent with the greater low-frequency content of typical
temporal power
spectra observed as fibrillation progresses
\cite{johnsonjacc,herbschleb80}.
Such restructuring  may be due to a coalescence of small
structures into larger
ones as time passes, perhaps related to the changing
parameters of the myocardium. Worley and colleagues noted decreased
activation rates in epicardial electrograms during the
first few minutes of VF in dogs; changes in activation rate on the
endocardium were much smaller
\cite{worley}. These effects were attributed to
the ischemic condition of the heart.
Numerical models have shown that even for fixed parameters, coalescence
of disorganized behavior into structures such as stable rotors can occur
\cite{markus}.

Can observed correlation lengths illuminate any aspect of the
relationship between low-dimensional chaos and fibrillation?
The correlation length of VF in pigs was found to vary between
4 and 10 mm.
Simple scaling arguments can be used to establish
a rough measure of the global complexity of fibrillation. If the correlation
length corresponds
to the length of the average coherent structure
on the epicardium, each such region would have an area of 16-100 mm$^{2}$.
The surface area of the ventricular epicardium in a typical pig heart
may be about 7500  mm$^{2}$. The ratio of these two areas implies that
at a given instant, there exist 75 to 500 approximately independent
areas of activity on the epicardium.
This rough estimate of the number of coherent regions, although probably
an upper bound (see Section~\ref{sec:lims}),
provides evidence that VF is too complicated
to be modeled as a low-dimensional, few-degree-of-freedom, system.
Such evidence could explain
the lack of convergence of the correlation dimension which has been
reported by
previous authors
\cite{kaplan+cohen,ravelli}.
Our results suggest that treating VF as low-dimensional chaos is unlikely
to succeed in models or in clinical strategies.

On the other hand these results strongly support the claim,
recently made in a study by Damle and co-workers
\cite{damle}, that VF in dogs is not random.
Damle et al. showed that the directions
of activation recorded by adjacent electrodes (separated by 2.5 mm)
were related. This is consistent with our observations
because the adjacent electrodes were
closer than a correlation length apart. Damle et al. also document
the decline in the amount of spatial ``linking'' over the first 5 seconds of
VF. They did not examine longer episodes of VF. The finite correlation
lengths we observed also provide direct evidence that VF is not a spatially
random process. Instead, both studies indicate that VF is a
high-dimensional process, but with a measurable degree of coherence.

\subsection{Limitations}
\label{sec:lims}

The estimate of 75 to 500 independent regions calculated above
might be interpreted as a generous {\em upper bound} on the
number of activation
fronts which exist simultaneously on the epicardium. The size of the electrode
array, which covers perhaps 5-10\% of the area of the ventricular epicardium,
prevents the observation of high correlations between events
further apart than the length of the array.
Thus, for behavior with a characteristic wavelength larger than
the size of the array, the estimate of correlation length
is biased downwards.
Also, our statistic
measures correlations between simultaneous events, so two sites
exhibiting similar activity
at different times will appear under-correlated. Finally, the choice of
the unscaled correlation length $\lambda$, rather than say $2 \lambda$,
to represent the length of a ``correlated'' region represents
a conservative and somewhat arbitrary decision.

Averaging over the area of the array and over one-second
blocks of time
may diminish the
contributions of local or short-lived organized processes.
However, the spatial correlation function is independent of
any temporal variability during the averaging time.
For example, as long as their spatial properties are the
same
a stationary spiral wave will have the same correlation
length as a drifting spiral, though their local temporal signals will
be quite different.
As another example,
if an identical, temporally random, signal were recorded by each electrode,
the electrodes would still be perfectly correlated.

No evidence has been presented of a direct relationship
between the correlation length and any other specific quantity
measured during VF.
Some of the conclusions
drawn in this paper are based on analogies
to other distributed
systems: notably fluid experiments and numerical simulations. Improved
experiments and methods of analysis may allow the calculation of
the mean distance between reentrant loops,
or a similar quantity, to firmly establish
the absolute meaning of the correlation length in fibrillation.
A possible length scale for comparison is the width of the region
of electromotive activity during fibrillation, which was recently
reported by Witkowski and colleagues
\cite{witkowski} to be between 0.5 and 1.0 mm in dogs; this
quantity is
roughly an order of magnitude
smaller than our estimated correlation lengths.

Although our measurements were taken from a 2-dimensional
array, VF occurs in a 3-dimensional
medium. It has been shown that well-defined 3-D structures can
produce apparently spatially complex patterns of activity on a
2-D surface
\cite{winfree80}. Therefore it is quite possible that even though
the 2-D patterns may be well characterized, the underlying 3-D process
is not.

\subsection{Implications}

It is potentially
extremely useful to measure quantities which
are associated with organization.
If the break-up or coalescence
of reentrant waves is part of the mechanism of fibrillation
or defibrillation, then a measure of the  coherence of VF
must be important to the clinician or the
electrophysiologist. Drugs may be compared by studying
whether, or how much,  they increase or decrease spatial organization.
Does more organization make it more or less difficult to
defibrillate? Does the optimal defibrillation
strategy depend to the spatial nature
of the arrhythmia? Is a high degree of organization
associated with a tendency to spontaneously defibrillate?
Is the size of a rotor related to the size of the smallest heart
that will sustain fibrillation?
It is possible to answer these questions only
if it is possible to compare unbiased measures of spatial structure.
The correlation length is a repeatable and
objective measure of relative organization. The description
of spatial organization in VF by such a quantity is a useful
step toward understanding and effectively treating fibrillation.

\section*{Acknowledgements}
We wish to acknowledge and thank Michael Yarger, Ed Simpson, and
Sharon Melnick for their considerable help, and D. Kaplan
for his helpful comments. Computational aspects of this work
were supported by a grant of computer time from the North Carolina
Supercomputer Center.

%\pagebreak

\pagebreak

\section*{Figures}

\noindent
Figure 1. \\
The spatial correlation function of artificial data which are periodic
in time and 2-dimensional space: $f(x,y,t)= \sin (4 t - 2 x - y)$.
a) The spatial pattern, $f(x,y, t_{0})$, observed at one instant in time
(a snapshot of the traveling wave).
b) The correlation, $R$, as a function of displacement, or
spatial lag, $[dx,dy]^{T}$. The data  consisted of
100 time samples at each ``sensor'' of a $15 \times 15$ spatial array.
\vspace{.15in}

\noindent
Figure 2. \\
Same as Figure 1. except that data which was generated
by a UNIX random number generator: $f(x,y,t) = $ {\tt random()}.
a) Random instantaneous spatial pattern. b) Correlation vs
vector displacement:
the spatial correlation is
approximately a delta-function (zero everywhere except at the origin).
\vspace{.15in}

\noindent
Figure 3. \\
a) A typical surface of correlation vs. separation vector (spatial lag).
b) The corresponding curve of correlation vs.
magnitude of sensor separation. Also shown are
the characteristic constants $\lambda$, $L_{e}$, and $L_{0}$.
Data was from 1000 samples of VF
at each sensor of a $22 \times 23$ array.
\vspace{.15in}

\noindent
Figure 4. \\
Photograph of the electrode array.
Silver/silver chloride electrodes
are arranged in a $22 \times 23$ grid with 1.12 mm inter-electrode
spacing.
\vspace{.15in}

\noindent
Figure 5. \\
A representative time series of VF recorded at
one electrode, one second after fibrillation
was induced by a premature (S2) stimulus.
The sampling frequency was 2000 samples/sec.
\vspace{.15in}

\noindent
Figure 6. \\
Spatial patterns of electrode potentials
recorded  by the $22 \times 23$ array during the
passage of an activation front 1 second after
the beginning of VF. Darkest pixels represent potentials of
less than -2 millivolts; lighter pixels represent progressively
higher potentials up
to +2 millivolts; potentials of
+2 millivolts and higher correspond to the lightest grey level.
(``Bad'' electrodes are responsible for
some isolated anomalous pixels). Panels represent ``snapshots'' of
the array  at
a) t=1.230 sec. b) t=1.250 sec. c) t=1.270 sec.
d) t=1.290 sec.
\vspace{.15in}

\noindent
Figure 7. \\
The spatial patterns of electrode potentials
recorded by the array 3 seconds after the
beginning of VF. a) t=3.210 sec. b) t=3.230 sec. c) t=3.250 sec.
d) t=3.270 sec.
\vspace{.15in}

\noindent
Figure 8. \\
Patterns of electrode potentials 1 minute after VF was induced.
a) t=61.690 sec. b) t=61.710 sec. c) t=61.730 sec.
d) t=61.750 sec.
\vspace{.15in}

\noindent
Figure 9. \\
Surfaces of spatial correlation of fibrillation vs
separation vector for experiment \#3.
Correlations were computed from
1.0 seconds of data at
each sensor of a $22 \times 23$ electrode array.
a) 1.0 seconds to 2.0 seconds after VF
was induced. b) 2.0 - 3.0 seconds. c) 3.0 - 4.0 seconds.
d) 60.0 - 61.0 seconds.
\vspace{.15in}

\noindent
Figure 10. \\
Curves of spatial correlation vs magnitude of separation for
experiment \#3. The radial distance from the
origin of Figures 8a-d) to each grid point
is the magnitude of separation. See Table 2 for the correlation
lengths corresponding to each curve.
\vspace{.15in}

\noindent
Figure 11. \\
Surfaces of spatial correlation vs separation vector for experiment \#4.
a) 1.0 - 2.0 seconds. b) 2.0 - 3.0 seconds. c) 3.0 - 4.0 seconds.
d) 60.0 - 61.0 seconds.
\vspace{.15in}

\noindent
Figure 12. \\
Curves of spatial correlation vs magnitude of separation for
experiment \#4. See Table 2 for the corresponding correlation lengths.
\vspace{.15in}

\noindent
Figure 13. \\
Mean ($\pm$ standard deviation) correlation lengths over the
five experiments at various
stages of VF. The difference
between correlation lengths at 60-61 seconds and those at 3-4 seconds
is statistically significant with $p < 0.05$.


\begin{thebibliography}{10}

\bibitem{wiggers}
C.J. Wiggers.
\newblock The mechanism and nature of ventricular defibrillation.
\newblock {\em American Heart Journal}, pages 399--412, 1940.

\bibitem{frazier}
D.W. Frazier, P.D. Wolf, J.M. Wharton, A.S.L. Tang, W.M. Smith, and R.E.
  Ideker.
\newblock Stimulus-induced critical point: Mechanism for electrical initiation
  of reentry in normal canine myocardium.
\newblock {\em J. Clinical Investigation}, pages 1039--1052, 1989.

\bibitem{allessie}
M.A. Allessie, F.I.M. Bonke, and F.J.G. Schopman.
\newblock Circus movement in rabbit atrial muscle as a mechanism of
  tachycardia. iii. the ``leading circle'' concept: A new model of circus
  movement in cardiac tissue without the involvement of an anatomical obstacle.
\newblock {\em Circulation Research}, pages 9--18, 1977.

\bibitem{ideker90}
R.E. Ideker, D.W. Frazier, W.~Krassowska, N.~Shibata, P-S. Chen, K.M. Kavanagh,
  and W.M. Smith.
\newblock Experimental eveidence for autowaves in the heart.
\newblock {\em Annals of the New York Academy of Sciences}, 519:208--218, 1990.

\bibitem{davidenko}
J.M. Davidenko, A.V. Pertsov, R.~Salomonsz, W.~Baxter, and J.~Jalife.
\newblock Stationary and drifting spiral waves of excitation in isolated
  cardiac muscle.
\newblock {\em Nature}, pages 349--351, 1992.

\bibitem{barkley}
Dwight Barkley.
\newblock A model for fast computer simulation of waves in excitable media.
\newblock {\em Physica D}, 49:61--70, 1991.

\bibitem{keener}
J.P. Keener.
\newblock A geometrical theory for spiral waves in excitable media.
\newblock {\em SIAM J. Appl. Math.}, pages 1039--1056, 1986.

\bibitem{gerhardt}
M.~Gerhardt, H.~Schuster, and J.J. Tyson.
\newblock A cellular automaton model of excitable media including curvature and
  dispersion.
\newblock {\em Science}, pages 1563--1566, 1990.

\bibitem{ideker81}
R.E. Ideker, G.J. Klein, L.~Harrison, W.M. Smith, J.~Kasell, K.~Reimer, A.G.
  Wallace, and J.J. Gallagher.
\newblock The transition to ventricular fibrillation induced by reperfusion
  after acute ischemia in the dog: a period of organized epicardial activation.
\newblock {\em Circulation}, 63(6):1371--1379, 1981.

\bibitem{elsherif}
N.~El-Sherif, R.~Mehar, W.B. Gough, and R.H. Zeiler.
\newblock Ventricular activation patterns of spontaneous and induced
  ventricular rhythms in canine one-day-old myocardial infarction: Evidence for
  focal and reentrant mechanisms.
\newblock {\em Circulation Research}, pages 152--166, 1982.

\bibitem{goldbergerphyd}
A.L. Goldberger, V.~Bhargava, B.J. West, and A.J. Mandell.
\newblock Some observations on the question: Is ventricular fibrillation
  `chaos'?
\newblock {\em Physica D}, 19:282--289, 1986.

\bibitem{kaplan+cohen}
D.~Kaplan and R.~Cohen.
\newblock Is fibrillation chaos?
\newblock {\em Circulation Research}, 67:886--892, 1990.

\bibitem{guevara+glass}
M.R. Guevara and L.~Glass.
\newblock Phase locking, period-doubling bifurcations, and irregular dynamics
  in periodically stimulated cardiac cells.
\newblock {\em Science}, pages 1350--1353, 1981.

\bibitem{holmesamr90}
P.J. Holmes.
\newblock Nonlinear dynamics, chaos, and mechanics.
\newblock {\em Applied Mechanics Review}, 43:S23--S29, 1990.

\bibitem{swinney+gollub}
H.L. Swinney and J.P. Gollub.
\newblock Characterization of hydrodynamic strange attractors.
\newblock {\em Physica D}, 18:448--454, 1986.

\bibitem{guckenheimer}
J.~Guckenheimer.
\newblock Dimension estimates for attractors.
\newblock {\em Contemporary Mathematics}, pages 357--367, 1984.

\bibitem{panchev}
S.~Panchev.
\newblock {\em Random Functions and Turbulence}.
\newblock Pergamon Press, New York, 1971.

\bibitem{lumley}
J.L. Lumley.
\newblock {\em Stochastic Tools In Turbulence}.
\newblock Academic Press, New York, 1970.

\bibitem{tufillaro}
N.B. Tufillaro, R.~Ramashankar, and J.P. Gollub.
\newblock Order-disorder transition in capillary waves.
\newblock {\em Physical Review Letters}, 62:422--425, 1989.

\bibitem{hohenberg+shraiman}
P.C. Hohenberg and B.I. Shraiman.
\newblock Chaotic behavior of an extended system.
\newblock {\em Physica D}, pages 109--115, 1989.

\bibitem{coullet}
P.~Coullet, L.~Gil, and J.~Lega.
\newblock Defect-mediated turbulence.
\newblock {\em Physical Review Letters}, pages 1619--1622, 1989.

\bibitem{panfilov}
A.V. Panfilov and A.V. Holden.
\newblock Spatiotemporal irregularity in a two-dimensional model of cardiac
  tissue.
\newblock {\em Int'l. J. Bifurcation and Chaos}, pages 219--225, 1991.

\bibitem{markus}
M.~Markus, M.~Krafczyk, and B.~Hess.
\newblock Randomized automata for isotropic modelling of two- and
  three-dimensional waves and spatiotemporal chaos in excitable media.
\newblock In A.V. Holden, M.~Markus, and H.G. Othmer, editors, {\em Nonlinear
  Wave Process In Excitable Media}, pages 161--182. Plenum Press, New York,
  1989.

\bibitem{pratt}
W.K. Pratt.
\newblock {\em Digital Image Processing}.
\newblock John Wiley \& Sons, New York, 1978.

\bibitem{baylycinc}
P.V. Bayly, E.E. Johnson, S.F. Idriss, R.E. Ideker, and W.M. Smith.
\newblock Minimum electrode spacing for mapping ventricular fibrillation using
  spatial sampling theory.
\newblock In {\em Proc. Computers in Cardiology}, Oct. 1992.

\bibitem{johnsoncirc}
E.E. Johnson, D.L. Rollins, P.D. Wolf, W.M. Smith, and R.E. Ideker.
\newblock Mechanism of ventricular fibrillation as mapped with 524 closely
  spaced simultaneously recorded epicardial electrodes.
\newblock {\em Circulation}, pages I--820, 1992.

\bibitem{patwolf}
P.D. Wolf, D.L. Rollins, T.F. Blitchington, R.E. Ideker, and W.M. Smith.
\newblock Design for a 512 channel cardiac mapping system.
\newblock In D.C. Mikulecky and A.M. Clarke, editors, {\em Biomedical
  Engineering: Opening New Doors. Proceedings of the Fall 1990 Annual Meeting
  of the Biomedical Engineering Society}, pages 5--13, New York, 1990. New York
  University Press.

\bibitem{johnsonjacc}
E.E. Johnson, S.F. Idriss, C.~Cabo, S.B. Melnick, W.M. Smith, and R.E. Ideker.
\newblock Evidence that organization increases during the first minute of
  ventricular fibrillation in pigs mapped with closely spaced electrodes.
\newblock {\em J. American College Cardiology}, page 90A, 1992.

\bibitem{moe}
G.K. Moe, W.C. Rheinbolt, and J.A. Abildskov.
\newblock A computer model of atrial fibrillation.
\newblock {\em American Heart Journal}, pages 200--220, 1964.

\bibitem{herbschleb80}
J.H. Herbschleb, R.M. Heethaar, I.~van~der Tweel, and F.L. Meijler.
\newblock Frequency analysis of the {ECG} before and during ventricular
  fibrillation.
\newblock In {\em Proc. Computers in Cardiology}, pages 365--368, 1980.

\bibitem{worley}
S.J. Worley, J.L. Swain, P.G. Colavita, W.M. Smith, and R.E. Ideker.
\newblock Development of an endocardial-epicardial gradient of activation rate
  during electrically induced, sustained, fibrillation in dogs.
\newblock {\em American J. Cardiology}, pages 813--820, 1984.

\bibitem{ravelli}
F.~Ravelli and R.~Antolini.
\newblock Dimensional analysis of the ventricular fibrillation {ECG}.
\newblock In A.V. Holden, M.~Markus, and H.G. Othmer, editors, {\em Nonlinear
  Wave Processes in Excitable Media}, pages 335--342. Plenum Press, 1989.

\bibitem{damle}
R.S. Damle, N.M. Kanaan, N.S. Robinson, Y.~Ge, J.J. Goldberger, and A.H.
  Kadish.
\newblock Spatial and temporal linking of epicardial activation patterns during
  ventricular fibrillation in dogs: evidence for underlying spatial
  organization.
\newblock {\em Circulation}, pages 1547--1558, 1992.

\bibitem{witkowski}
F.Witkowski, P.A. Penkoske, and K.M. Kavanagh.
\newblock Estimate of electromotive surface dimension during ventricular
  fibrillation.
\newblock In {\em Proc. IEEE Engineering in Medicine and Biology Society
  Conference}, pages 32--33, 1991.

\bibitem{winfree80}
A.T. Winfree.
\newblock {\em The Geometry of Biological Time}, chapter~9.
\newblock Springer-Verlag, New York, 1980.

\end{thebibliography}
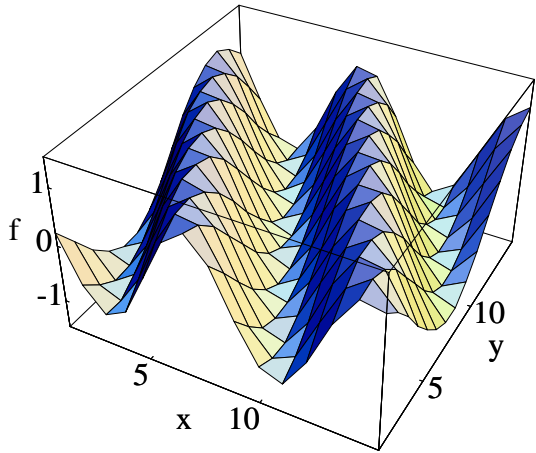
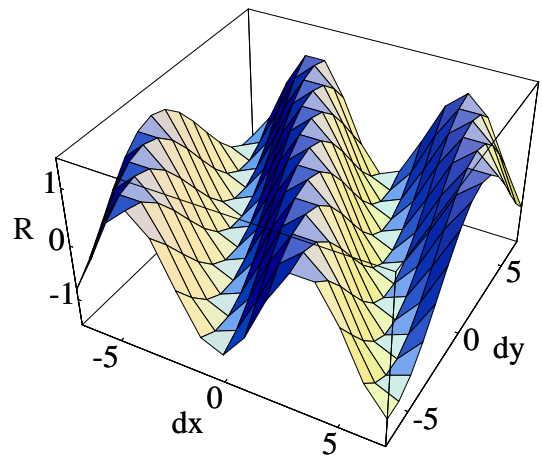
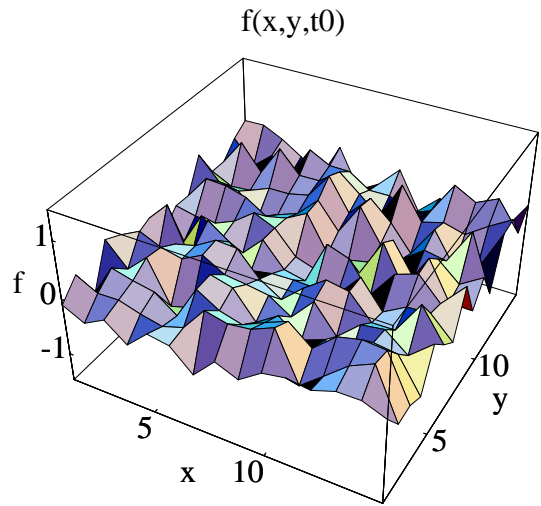
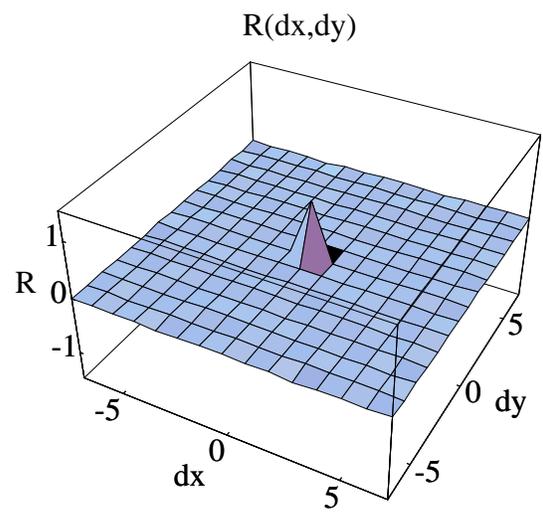
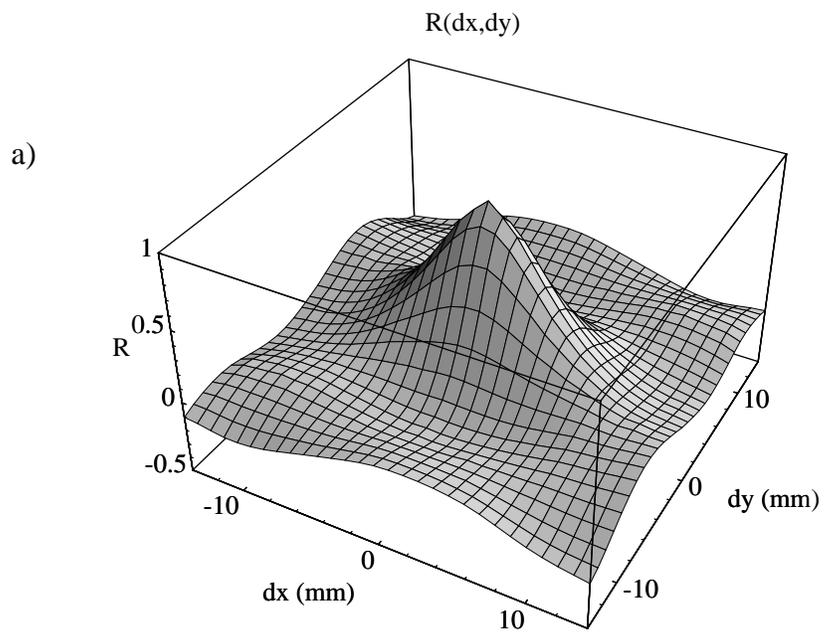
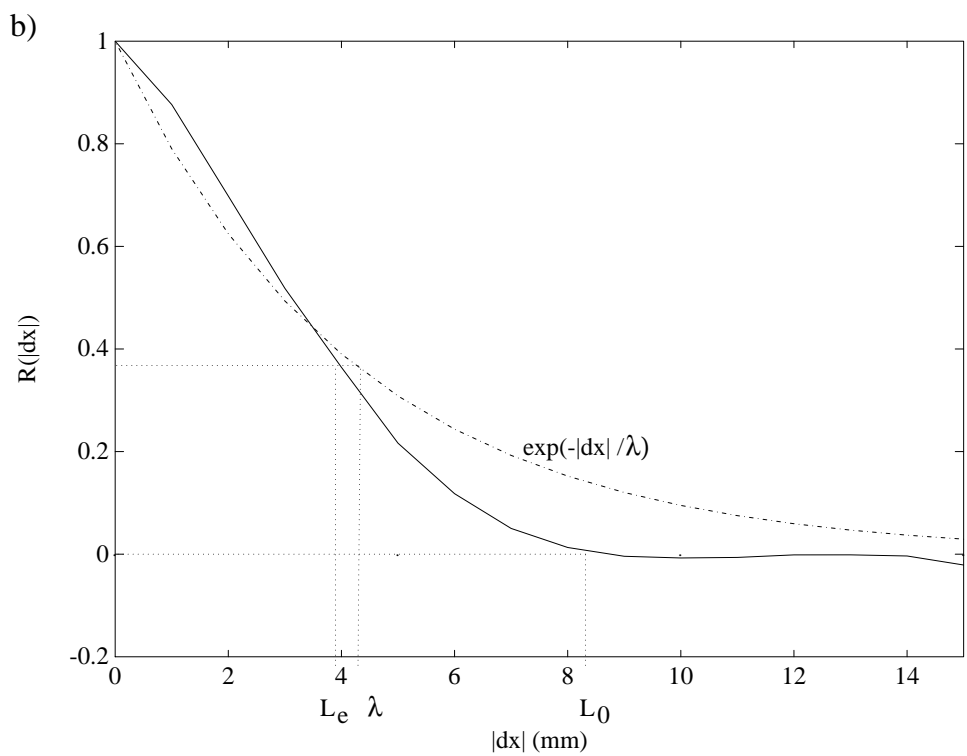
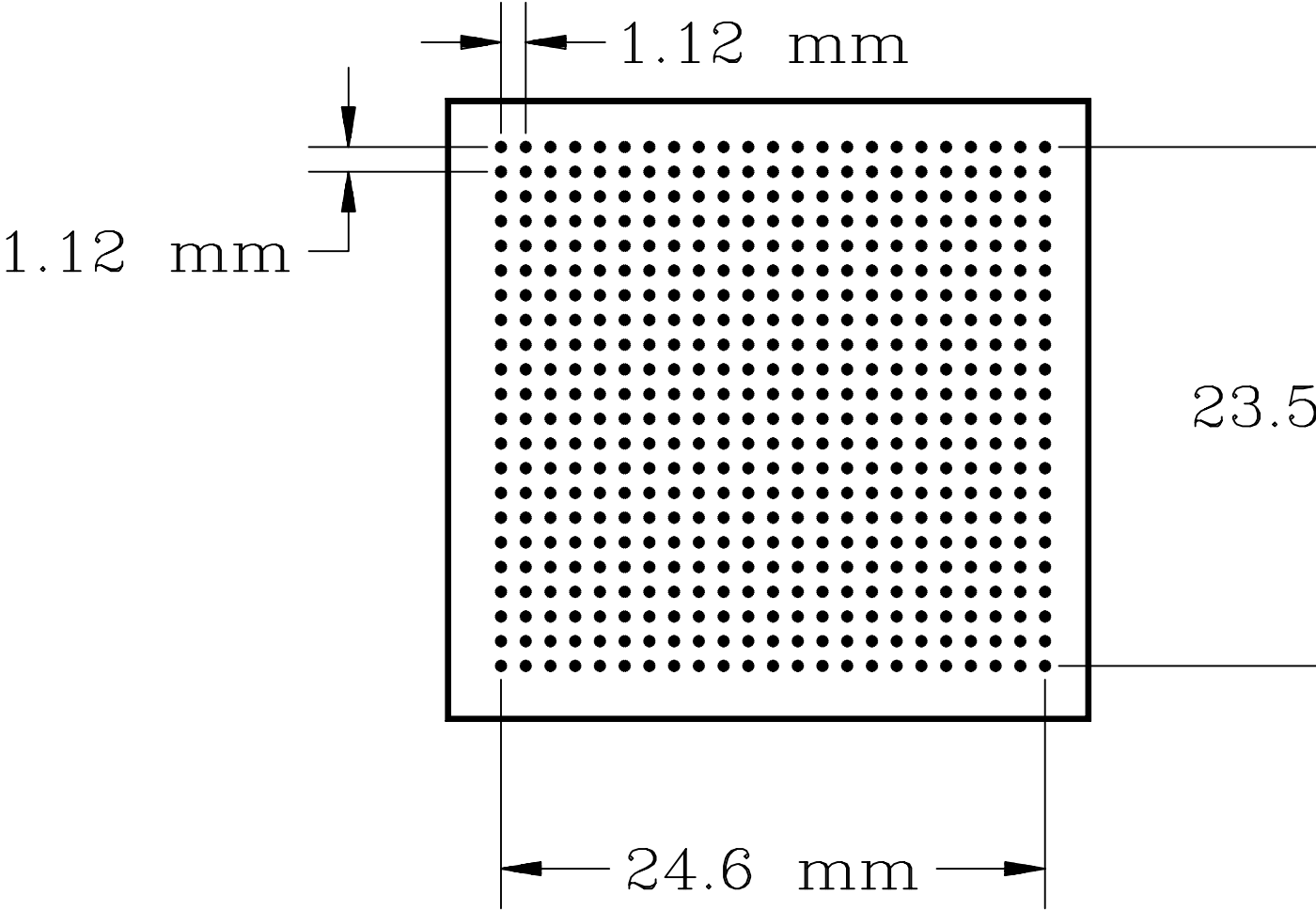
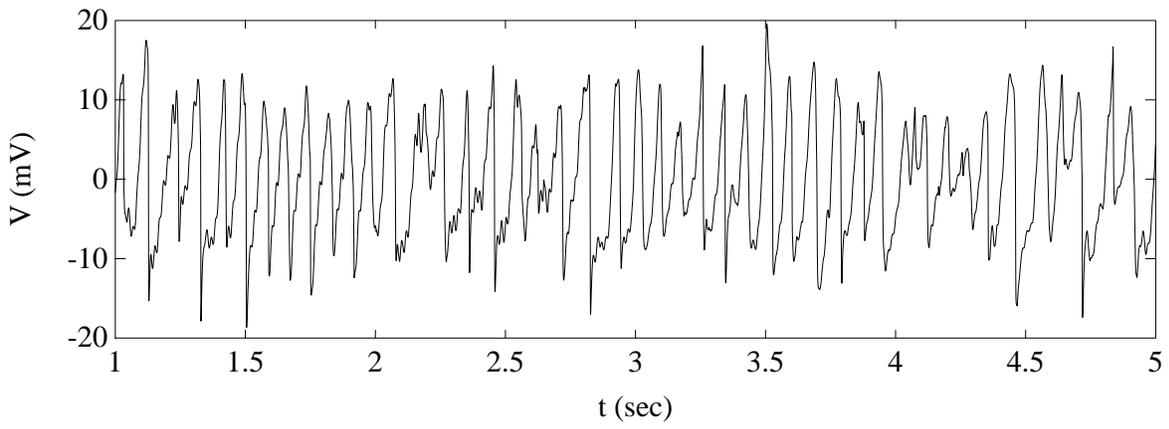
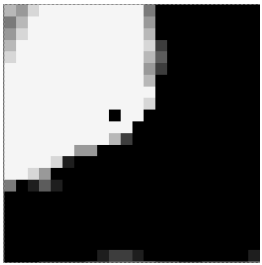
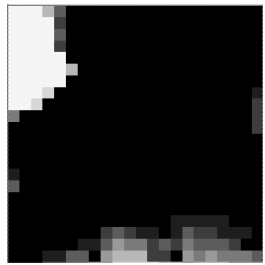
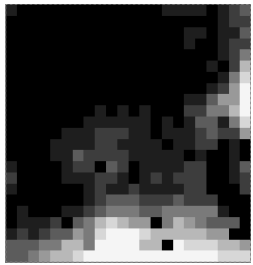
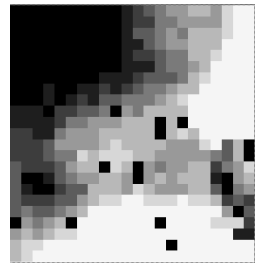
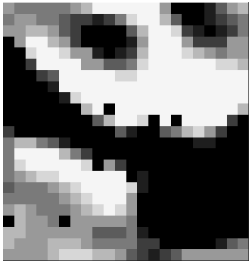
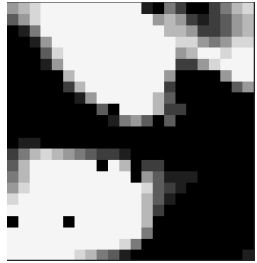
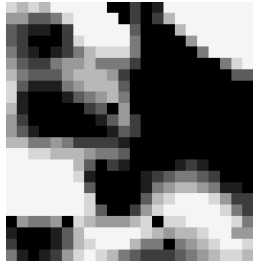
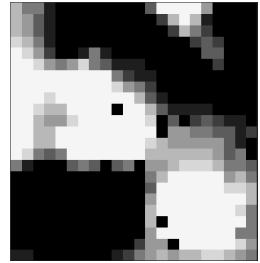
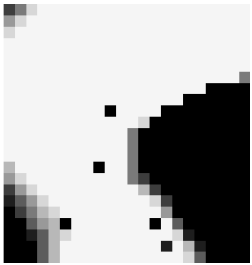
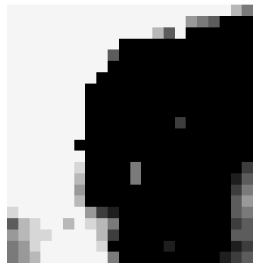
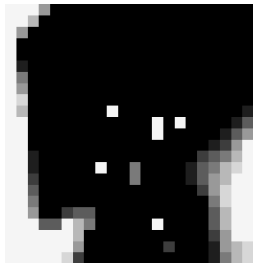
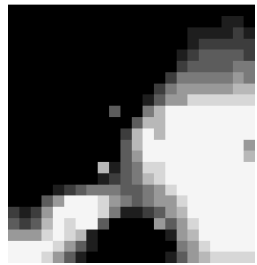
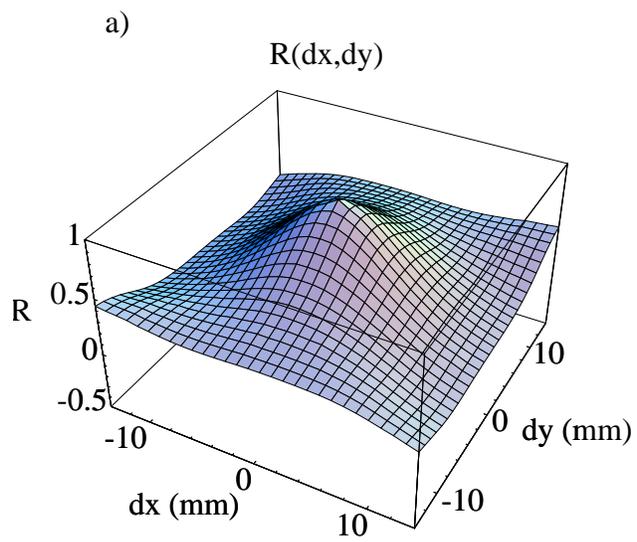
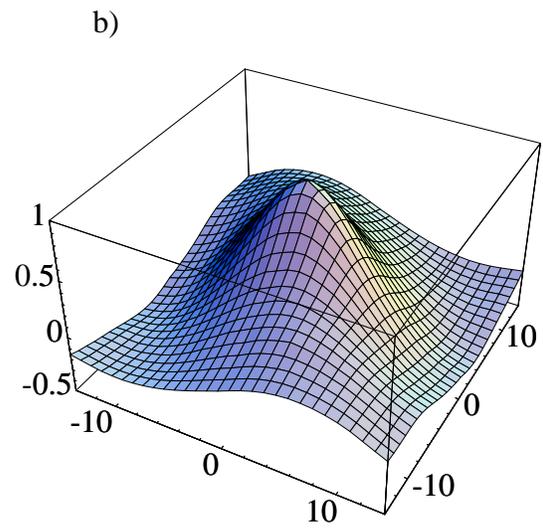
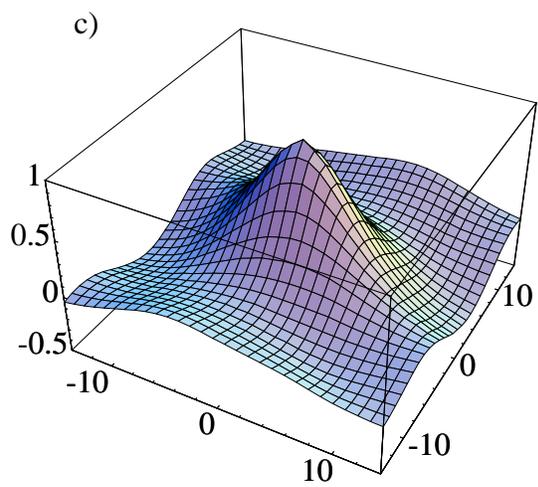
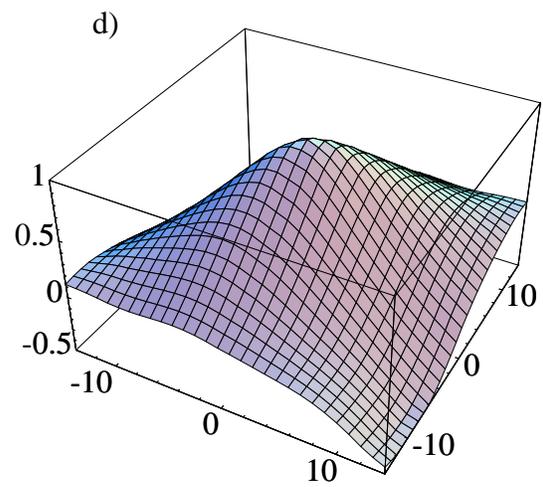
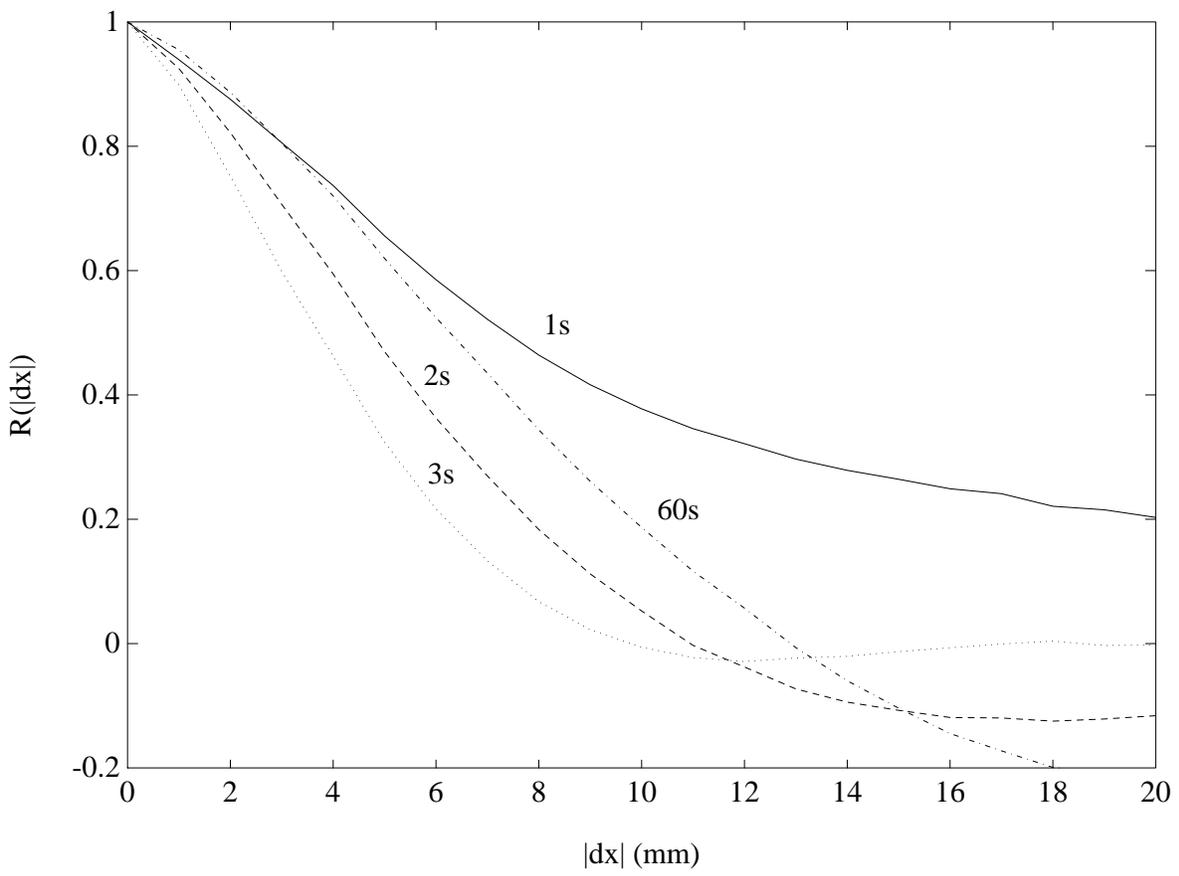
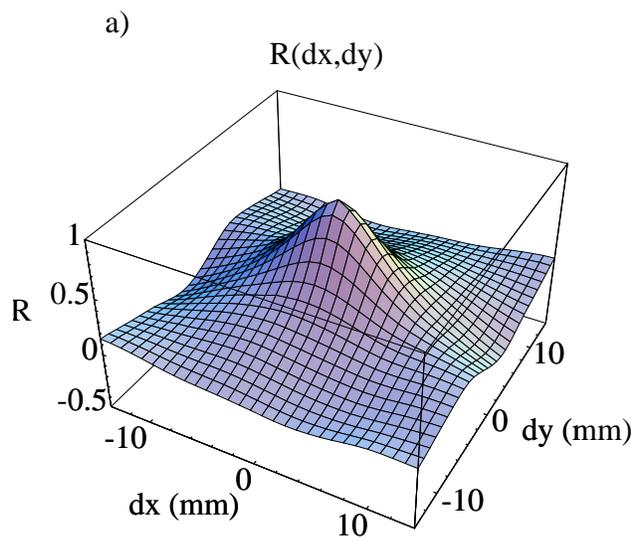
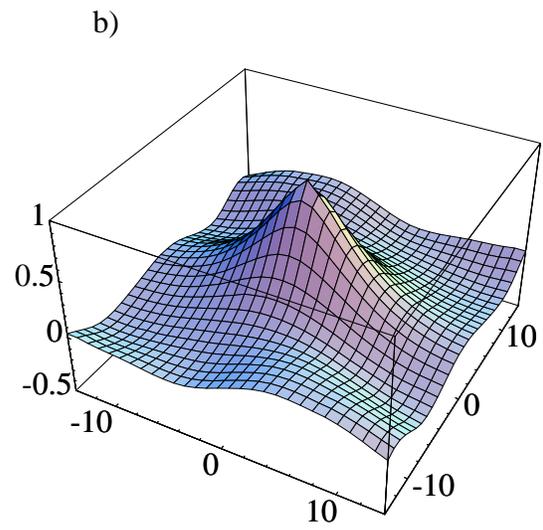
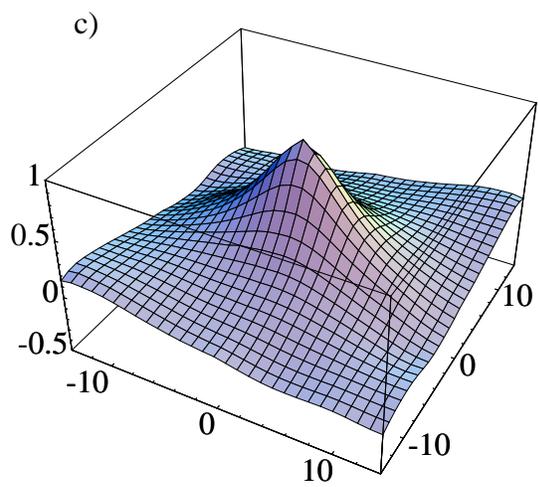
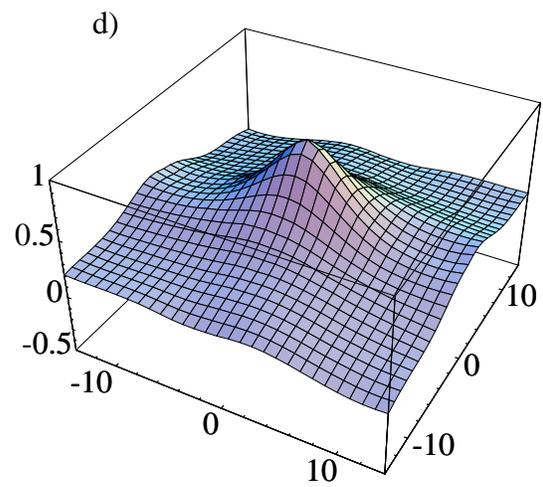
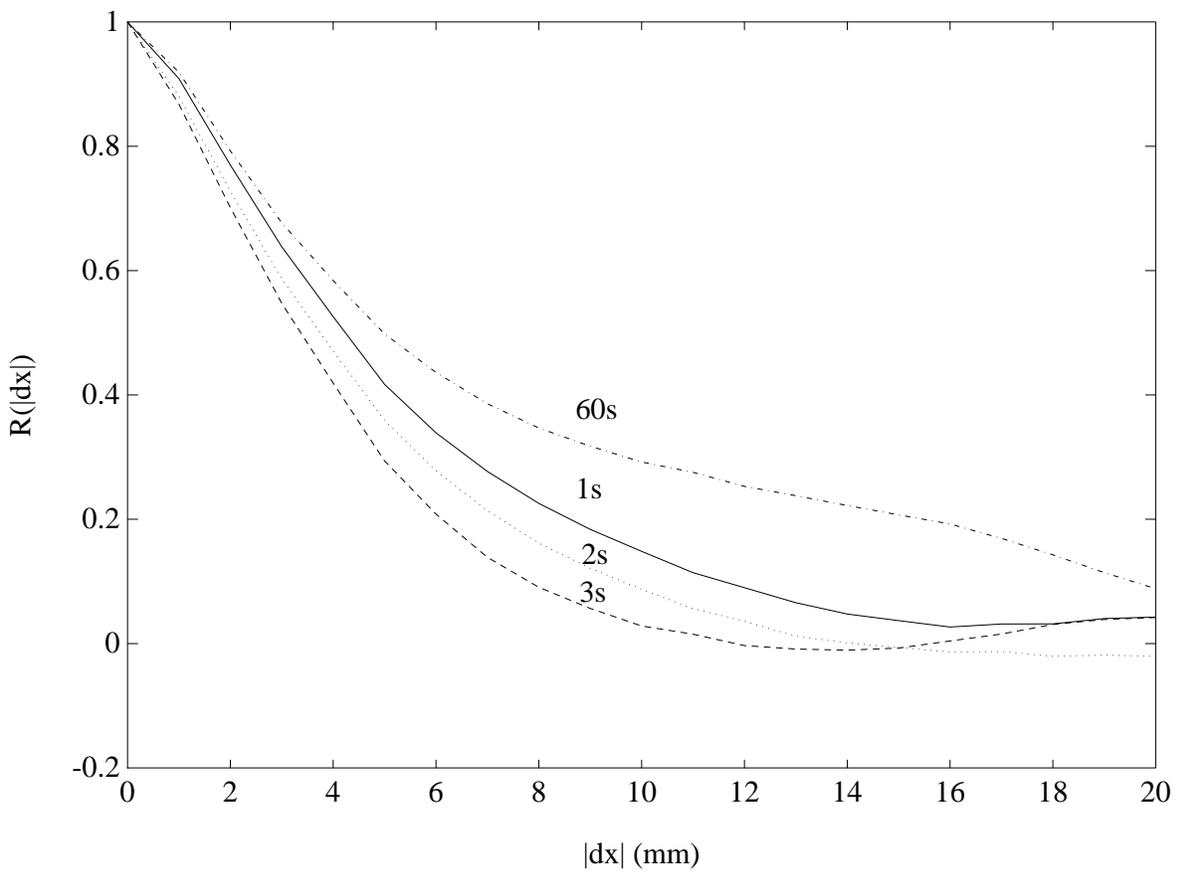
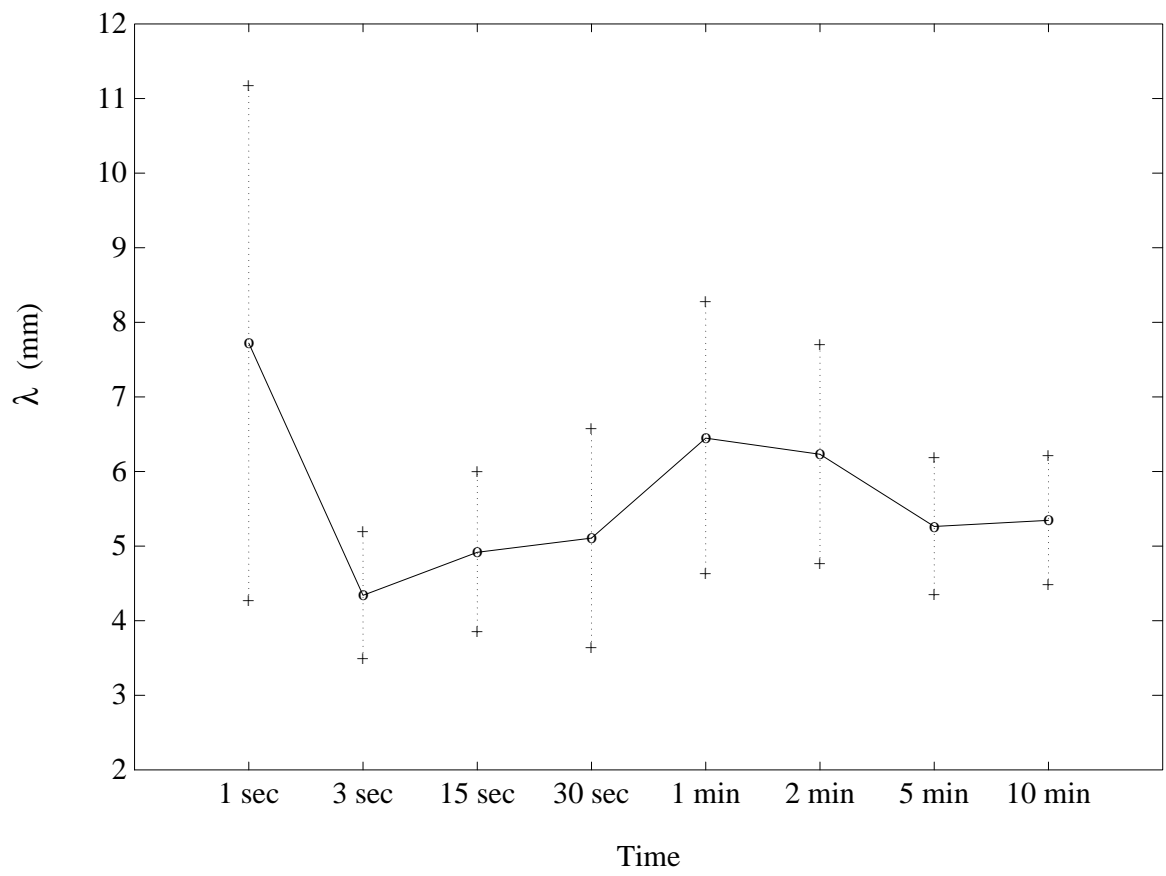
\end{document}